\documentclass[9pt,twocolumn,twoside]{opticajnl}
\journal{opticajournal}

\usepackage{subfig}
\usepackage[percent]{overpic}
\usepackage{xcolor}
\usepackage{textcomp}

\setboolean{shortarticle}{true}

\usepackage{lineno}
\usepackage{siunitx}
\usepackage{textalpha}
\usepackage{upgreek}
\usepackage{bm, upgreek}

\title{Octave-spanning 10-GHz Er-doped solid-state optical frequency comb}

\author[1,\dag]{Niccolò S. Barberio}
\author[2,\dag,*]{Francesco Canella}
\author[3]{Andrea Pertoldi}
\author[3]{Benjamin Rudin}
\author[3]{Oguzhan Kara}
\author[3]{Florian Emaury}
\author[1]{Antonio Caruso}
\author[1]{Dario Giannotti}
\author[4,5]{Francesco Leone}
\author[1,5]{Paolo Laporta}
\author[2,5,**]{Gianluca Galzerano}

\affil[1]{Dipartimento di Fisica - Politecnico di Milano, Piazza Leonardo Da Vinci, 32, 20133 Milano, Italy}
\affil[2]{Istituto di Fotonica e Nanotecnologie - Consiglio Nazionale delle Ricerche, Piazza Leonardo Da Vinci, 32, 20133 Milano, Italy }
\affil[3]{Menhir Photonics AG, Zürichstrasse 130, 8600 Dübendorf, Switzerlan}
\affil[4]{Dipartimento di Fisica e Astronomia, Sezione Astrofisica, Università di Catania, Via Santa Sofia, 64
95123 Catania, Italy}
\affil[5]{INAF–Osservatorio Astrofisico di Catania, Via Santa Sofia 78, 95123 Catania, Italy}

\affil[$\dag$]{These two authors equally contributed to this work}
\affil[*]{francesco.canella@cnr.it}
\affil[**]{gianluca.galzerano@cnr.it}

\begin{abstract}
Optical frequency combs provide a phase-coherent interface between optical and microwave domains, underpinning advances in precision metrology, spectroscopy, and time–frequency transfer.
Most conventional comb architectures are limited to sub-gigahertz repetition rates, constraining integration and scalability.
Here, we demonstrate a compact, 10-GHz Er-doped solid-state frequency comb operating near the 1.55 $\bm{\upmu}$m telecommunications window. The system, assembled entirely from commercially available components, produces a coherent spectrum spanning from 1150~nm to 2350~nm with exceptionally low intensity and phase noise.
Full frequency stabilization of both pulse repetition and carrier envelope offset frequencies is demonstrated with respect to an RF reference.
Comprehensive characterization reveals performance exceeding that of previously reported solid-state combs in power-per-mode and noise suppression. These results establish a robust and accessible platform for multi-gigahertz-repetition-rate comb synthesis, bridging laboratory-grade performance with practical deployment in optical communication, precision timing, and astronomical instrumentation.
\end{abstract}

\setboolean{displaycopyright}{false} 

\begin{document}

\maketitle

\section{Introduction}
Optical frequency combs have revolutionized precision metrology, spectroscopy, and time–frequency transfer by providing a phase-coherent link between optical and RF/microwave domains \cite{Jones_2000,Udem_2002,Fortier2019}.
Conventional fiber-based combs (especially Er- and Yb-doped fiber combs) currently represent the most mature and widely validated platform for achieving the highest levels of coherence and absolute frequency accuracy \cite{Kim_16,Hansel_2018}. However, their relatively low repetition rates (in general $<1$~GHz) pose challenges for integration and widespread deployment. Alternative solid-state comb technologies such as microresonator (Kerr) combs \cite{Delhaye_2007,Kippenberg_2011,Herr_2014}, electro-optic \cite{Kobayashi_1972,Kourogi_1993,Parriaux_20} and semiconductor combs \cite{Moskalenko_14,Duan_2014,Wang_2017}, are rapidly advancing and narrowing the performance gap.
Solid-state frequency combs operating near the 1.55~$\upmu$m telecommunications window with mode spacings around tens of GHz represent a compelling evolution of this technology.
Indeed, a 10-GHz comb spacing aligns naturally with dense wavelength-division multiplexing grids, supporting next-generation optical communication systems that exploit mutually coherent carriers across the C-band \cite{Pfeifle_2014,Ataie_2015,Marin_2017}.
In parallel, such combs serve as precise optical frequency dividers for the synthesis of ultralow-phase-noise microwave signals \cite{Millo_2009,Fortier_2011,Torres2014}, and as stable, evenly spaced spectral references for the calibration of near-infrared astronomical spectrographs \cite{Steinmetz_2008,McCracken_2017}.

Among the various solid-state frequency comb platforms reported above, such as microresonator Kerr combs, semiconductor mode-locked lasers, and electro-optic combs \cite{Krainer_2002,Schlatter_05,Bartels2009,Kimura2019, Sekhar_2023}, which all support high repetition rates (up to hundreds GHz) and broad spectral coverage, optical frequency combs synthesized by Er-doped mode-locked lasers have demonstrated higher power-per-mode performance (exceeding mW level) and remarkably low phase noise and time jitter \cite{Stumpf_2010,Schibli_2016,Lesko_2020}.
Notably, the latter -pulse timing jitter-, scales quadratically with the fundamental pulse repetition rate, in sharp contrasts to the trend toward increasing repetition rates \cite{Pachotta_2004}.

\begin{figure*}[h]
\centering
\includegraphics[width=0.65\textwidth]{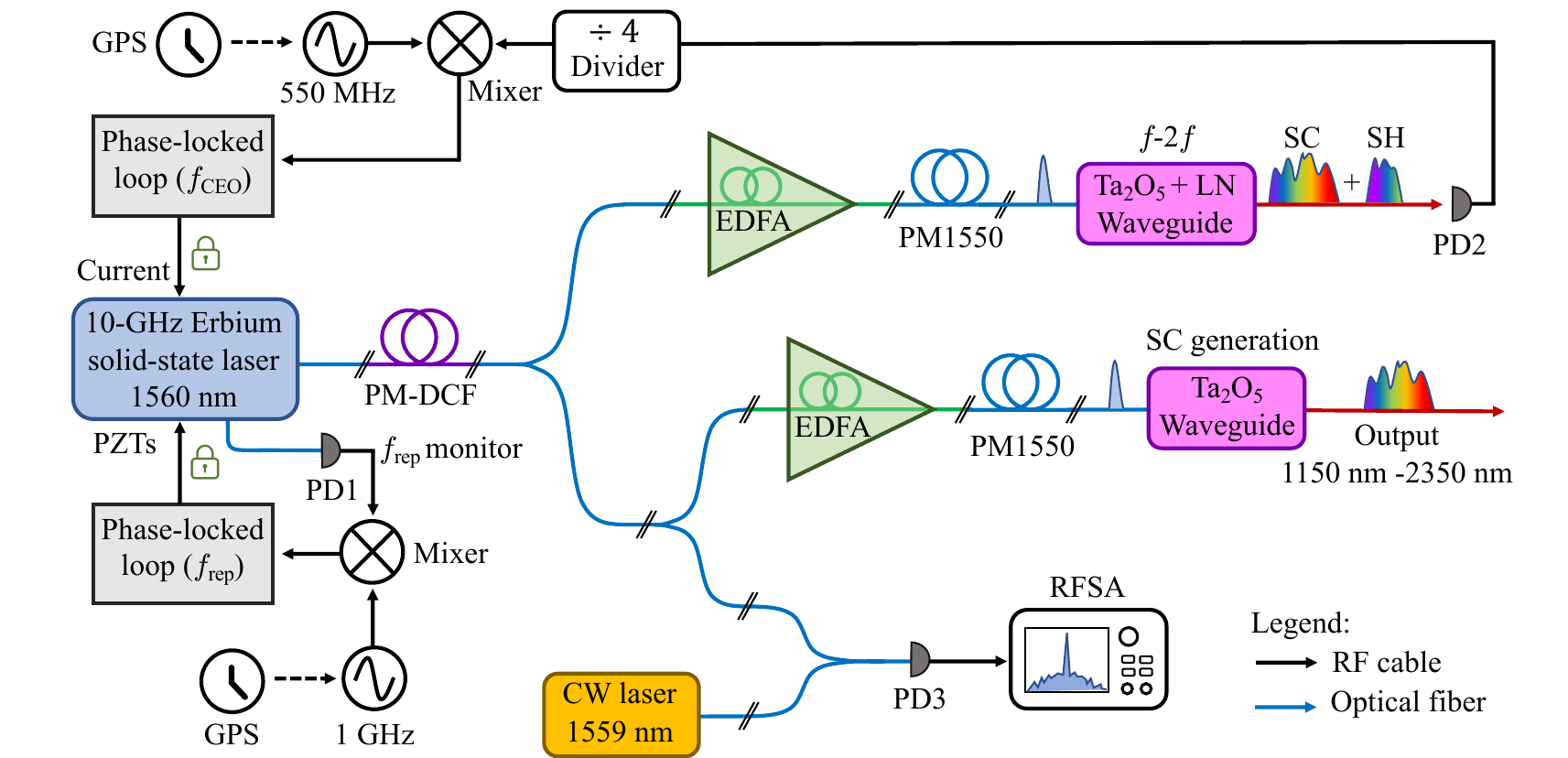}
\caption{Simplified experimental setup of the 10-GHz optical frequency comb. 
CW laser: single-frequency laser; PM-DCF: polarization-maintaining dispersion compensated  fiber; PM-1550: single-mode polarization-maintaining fiber; EDFA: erbium-doped fiber amplifier; PD: photodetector; PZT: piezoelectric transducer; GPS: GPS-disciplined Rb frequency standard; RFSA: RF spectrum analyzer; SC: supercontinuum; SH: second harmonic.}
\label{setup}
\end{figure*}

Here, we report on the realization and comprehensive characterization of a 10-GHz Er-doped solid-state optical frequency comb exhibiting exceptionally low intensity and phase noise, covering a spectral bandwidth spanning from 1150~nm to 2350~nm.
The system, entirely assembled from commercially available all-fiber coupled components, demonstrated the feasibility of a fully frequency stabilized 10~GHz optical frequency comb with high-performance operation. Both pulse repetition rate $f_{\text{rep}}$ and carrier envelope offset frequency $f_{\text{CEO}}$ are stabilized against a Rb frequency reference and GPS. The full characterization of the combs is obtained through measurements of the power spectral densities of intensity and phase noise, together with the optical beat note against a narrow-linewidth, single-frequency CW laser at 1560~nm. The $f_{\text{CEO}}$ signal produced at the output of an integrated $f$-$2f$ nonlinear interferometer exhibits a 50~dB signal-to-noise ratio within a 300~kHz measurement bandwidth, along with a sub-50~kHz linewidth over 0.1~ms observation times in free-running operation. A comb optical linewidth narrower than 10~kHz over 0.1~ms observation times is measured at $\sim$1560~nm.
The combination of compactness, high repetition rate, and high-performance operation positions this technology to extend the applicability of Er-doped solid-state frequency combs beyond laboratory environments, opening new opportunities in optical communication, precision timing, and astronomical instrumentation \cite{Pfeifle_2014,Ataie_2015,Marin_2017,Millo_2009,Fortier_2011,Torres2014,Steinmetz_2008,McCracken_2017}.

\section{Experimental setup}
\begin{figure}[h]
    \centering
    \includegraphics[width=\columnwidth]{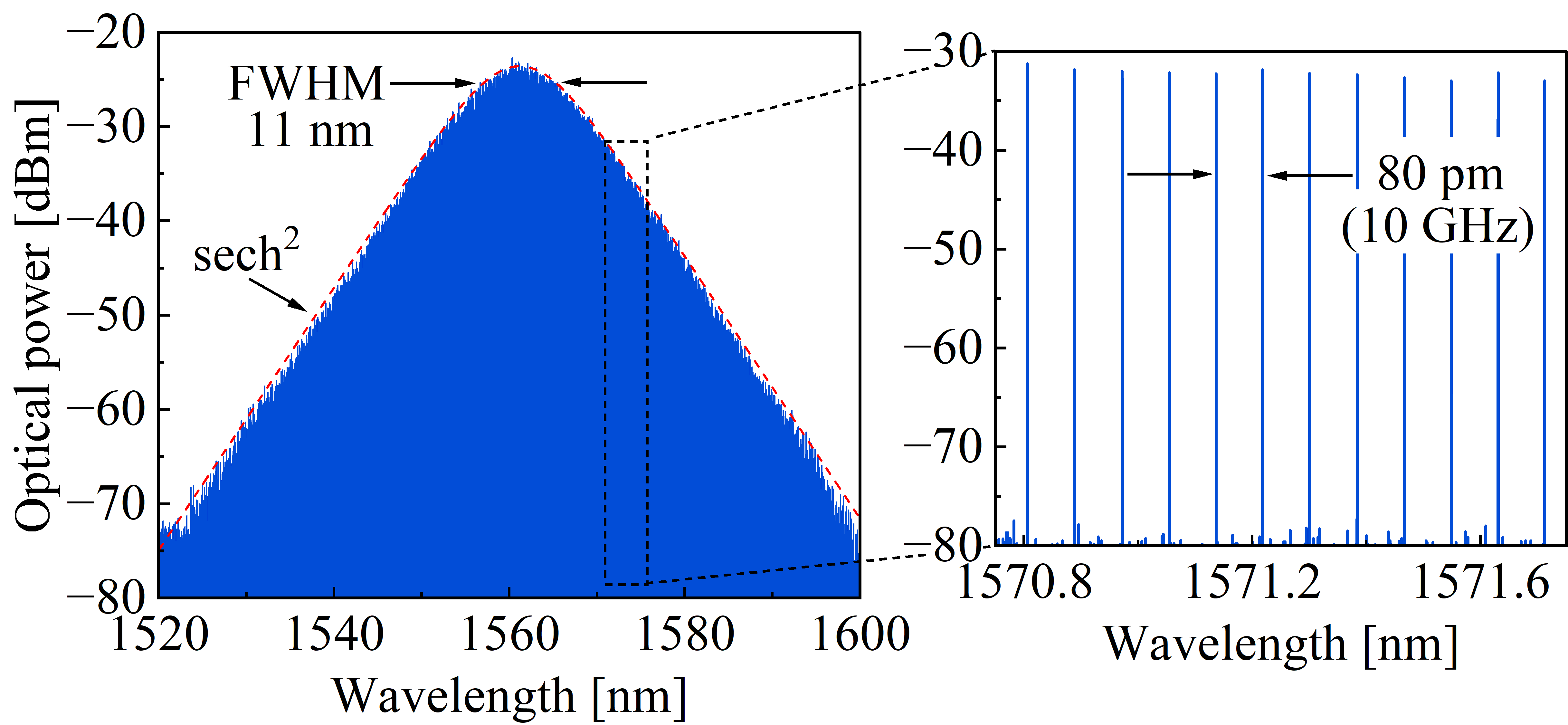}
    \caption{Optical spectrum and zoom showing the mode separation of 80~pm, corresponding to 10~GHz in frequency. The red dashed curve is the hyperbolic secant best fit profile.}
    \label{RFspectrum}
\end{figure}

Figure \ref{setup} shows the implemented setup of the 10~GHz optical frequency comb.
A passively mode-locked Er-doped solid-state laser (Menhir-1550 10 GHz from Menhir Photonics) generates an ultrafast pulse train at 10~GHz repetition rate with an average output power of 35~mW and a transform-limited pulse duration of $\sim$200~fs, corresponding to an optical bandwidth of 11~nm around the 1560~nm central emission wavelength. 
Figure \ref{RFspectrum} shows the optical spectrum measured using a high-resolution optical spectrum analyzer (Apex Technologies OSA-AP6) at the laser output together with the hyperbolic secant best interpolation profile.

\begin{figure}[h]
    \centering
    \includegraphics[width=\columnwidth]{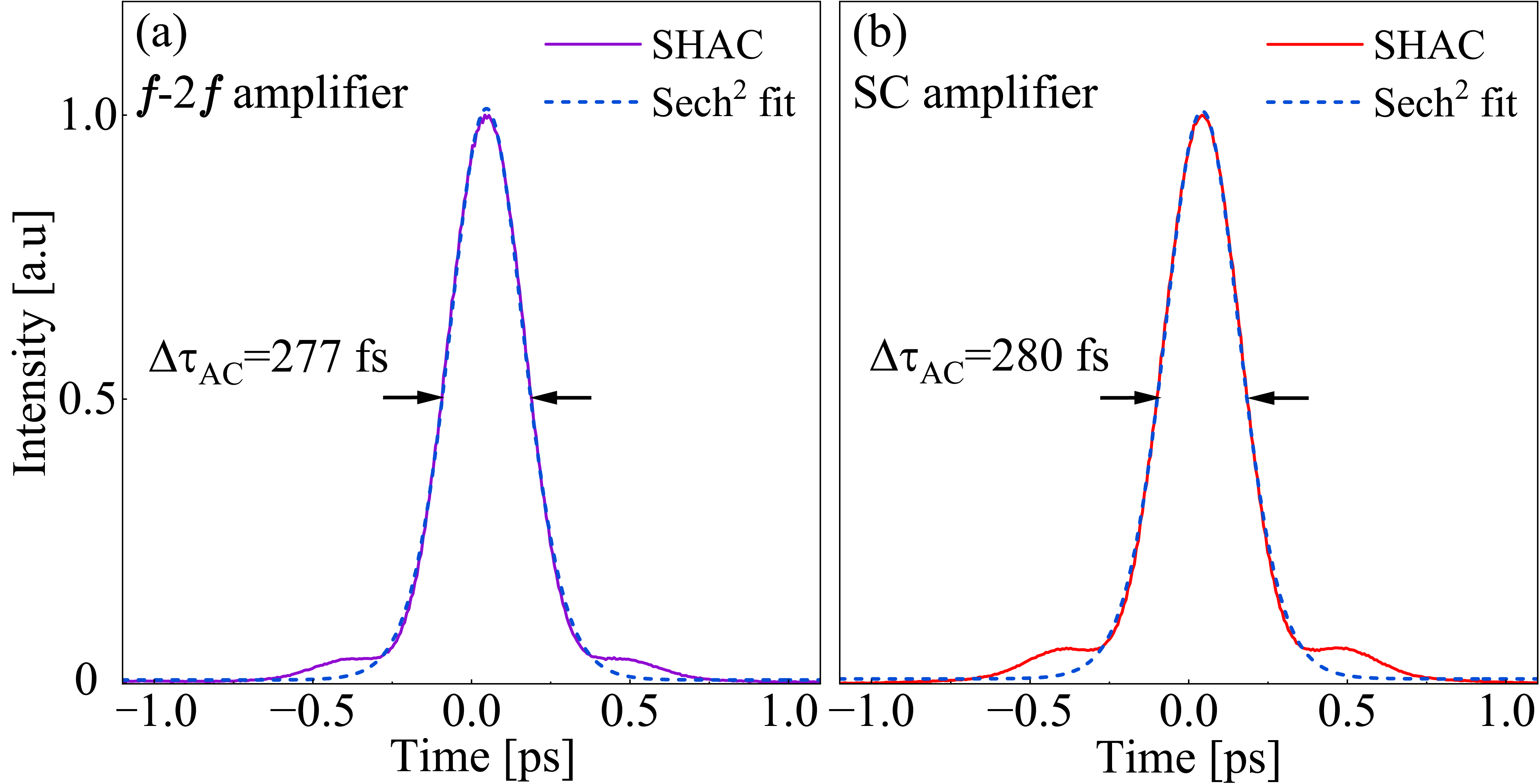}
    \includegraphics[width=\columnwidth]{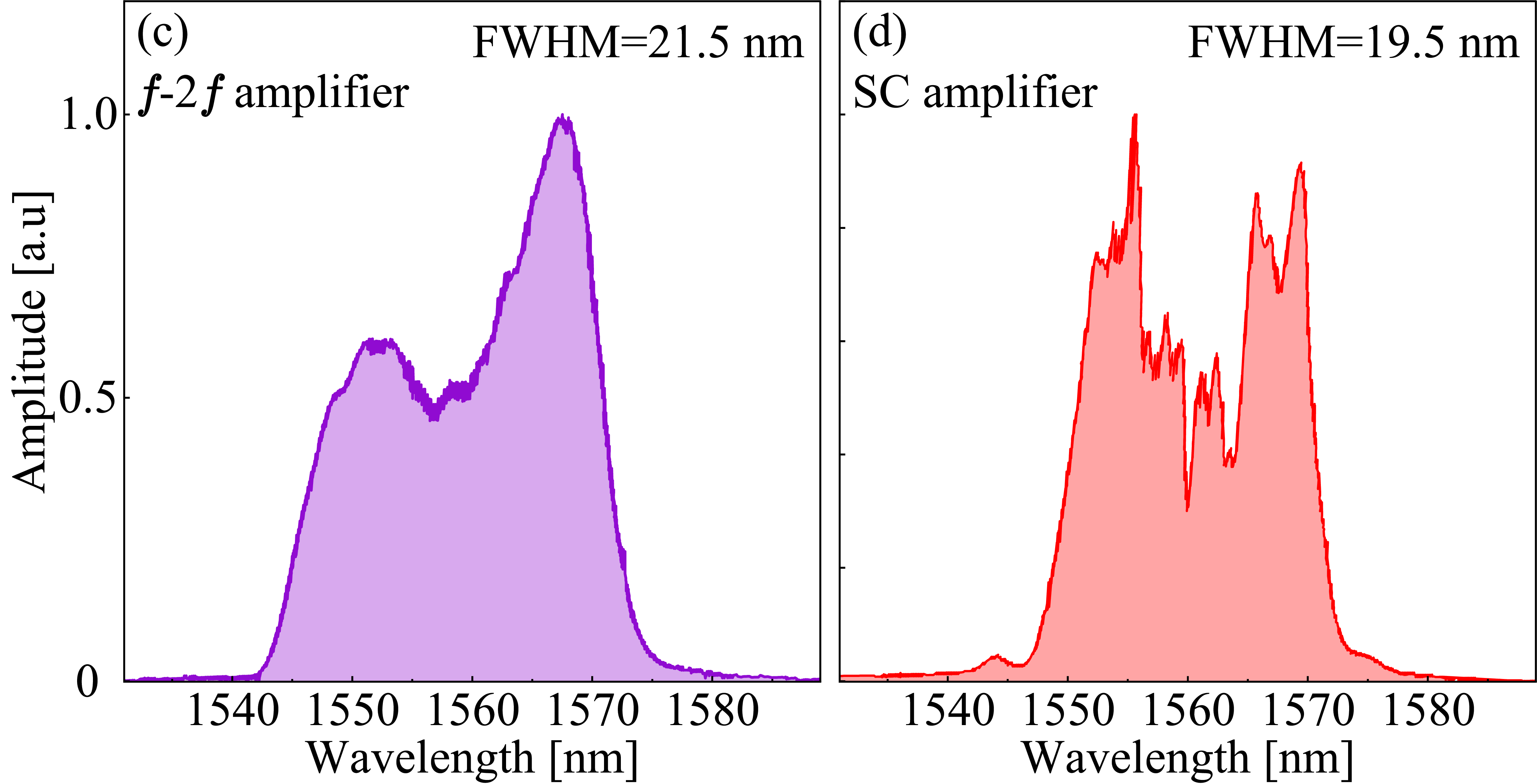}
    \caption{Autocorrelation traces and optical spectra measured at the outputs (a)-(c) of the $f-2f$ amplifier and (b)-(d) SC amplifier.}
    \label{ACandSpectra}
\end{figure}

The 10-GHz laser source after propagating through a 3.5~m long polarization-maintaining dispersion compensated fiber (PM-DCF, from Cycle) is then split into two different chirped pulse amplification (CPA) stages: the first one is devoted to the detection of the carrier envelope offset frequency ($f_{\text{CEO}}$) whereas the second one is used for the generation of a broadband supercontinuum (SC).
More specifically, the $f_{\text{CEO}}$ branch is constituted by an erbium-doped fiber amplifier (EDFA) with a maximum output power of 2.2~W (Nuphoton, EDFA-1560-PM-RS-33-23-FCA), a 2.3~m long PM-1550 single-mode fiber for pulse compression and $f-2f$ interferometer. The interferometer consists in a customized tantala nonlinear waveguide followed by a lithium niobate crystal in the same integrated device (Octave Photonics COSMO-O) for the simultaneous generation of a dispersion shifted SC and second-harmonic (SH) of the pulse train at around 780~nm. Such an integrated device, in combination with three optical filters (dichroic, low-pass and bandpass) to select only the radiation around 780~nm with a bandwidth up to 10~nm, and a free-space fast GaAs photodetector (Coherent GaAs PIN detector ET4000EXT) is used for the detection of the $f_{\text{CEO}}$ beatnote. The lowest-frequency carrier envelope offset alias at 2.2~GHz ($f_{rep}- f_{\text{CEO}}$) is amplified (two Mini-Circuits ZX60-153LN-S+), filtered (Mini-Circuits VLF-4400+), and divided by a factor of four by means of a frequency divider (Pasternack PE88D4003). The resulting signal is phase-referenced to a quartz oscillator using a digital proportional–integral–derivative (PID) servo (TEM Digital PhaseLock), which actuates the laser via the pump diode current to stabilize $f_{\text CEO}$.
The SC branch is made of two EDFA amplifiers in series to reach up to 4 W output power (Nuphoton EDFA-1560-PM-RS-23-13-FCA as preamplifier at 200~mW, and BKtel 4 W x2:THPOA-1.5-SP400a-FCAPC-FCAPC) with an additional 2.3~m long DCF-PM input fiber, a 2.6~m long PM-1550 single-mode fiber for pulse compression, and a tantala nonlinear waveguide (Octave Photonics DESMO) for the generation of a broadband SC. For simplicity, the preamplifier and the related 2.3~m DCF are not drawn in Fig.\ref{setup}.
In addition, for heterodyne characterization of the 10-GHz laser and absolute measurement of the carrier envelope offset frequency, 1\% of the EDFA output is combined with a narrow-linewidth, single-frequency laser at 1559.513~nm (TeraXion LXM).
The pulse repetition rate, measured with a fiber coupled fast InGaAs photodetector (Coherent ET-3500FEXT), is also referenced to a quartz oscillator via digital proportional–integral–derivative (PID) servos (TEM Digital PhaseLock) acting on the two laser's piezoelectric transducers. Both $f_{\text{CEO}}$ and $f_{\text rep}$ can be stabilized to a GPS-disciplined rubidium frequency standard.

\section{Results}
For efficient supercontinuum (SC) generation and $f_{CEO}$ detection, it is necessary to pump the tantala waveguides with pulses with a peak power exceeding 1~kW, corresponding to a 200~pJ in energy and a compressed duration of about 200~fs. Figure~\ref{ACandSpectra} shows the second-harmonic intensity autocorrelation (SHAC) traces and corresponding optical spectra of the compressed pulses measured at the output of the PM-1550 fibers following both the EDFAs. At the maximum output power levels, SHAC traces exhibit full widths at half maximum (FWHMs), respectively, of 277~fs and 280~fs for the $f-2f$ amplifier and for the SC amplifier. Assuming a Gaussian deconvolution factor, the corresponding pulse duration would be $201$~fs and $203$~fs, respectively. The pulse durations are verified against the transform-limited durations calculated from the EDFAs' optical spectra (Figure~\ref{ACandSpectra}(c) and (d)), where self-phase modulation-induced spectral broadening is about 20~nm FWHM for both plots, resulting in $210$~fs and $222$~fs.
The maximum power per comb line is estimated to exceed 9~mW for the $f_{\text{CEO}}$ amplifier and 11~mW for the SC amplifier.

\begin{figure}[t]
    \centering
    \includegraphics[width=\columnwidth]{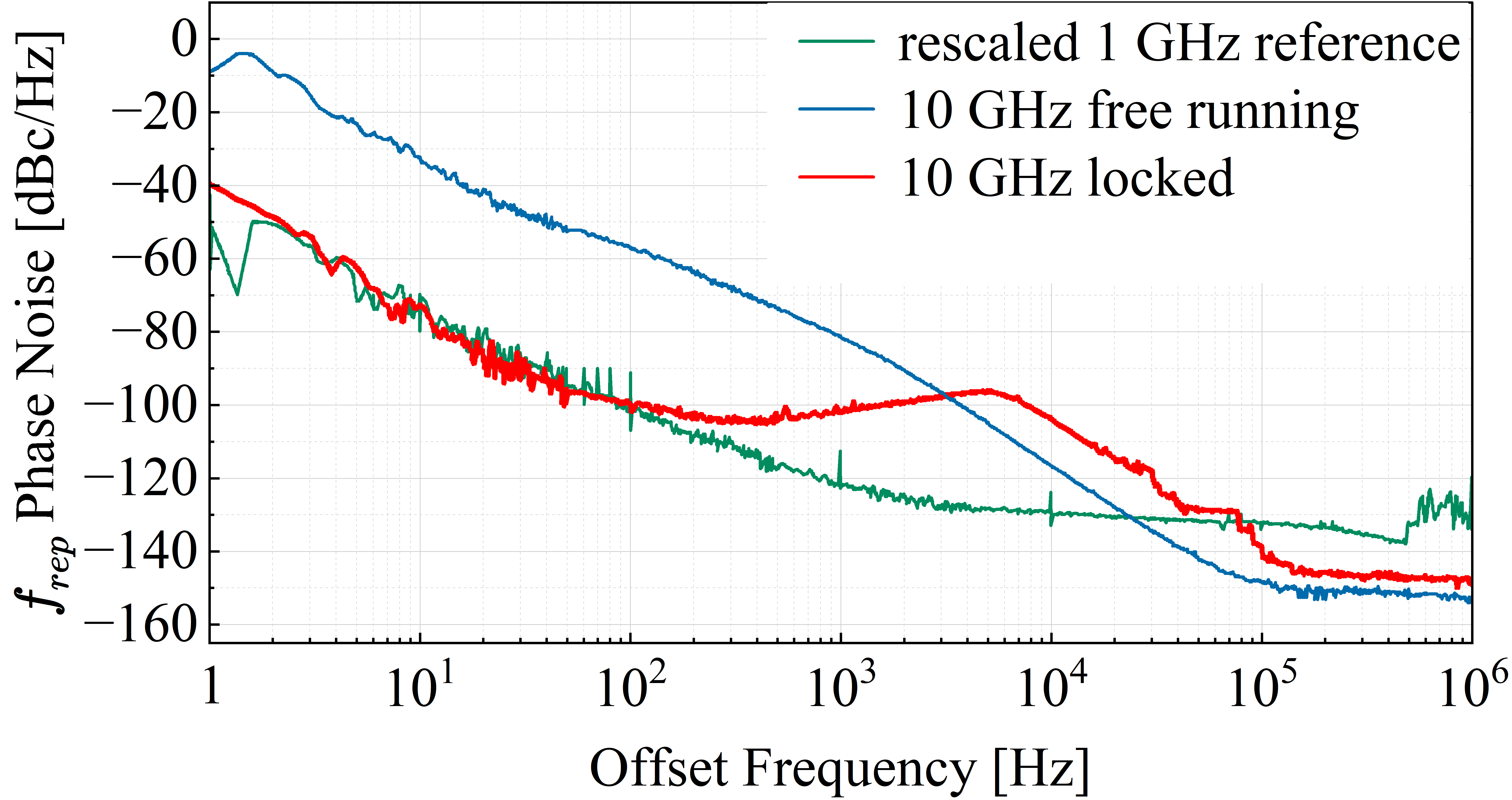}
    \caption{Phase noise of the 10-GHz pulse repetition rate both in free-running (blue trace) and locked (red trace) to the 1-GHz reference (dark green trace, rescaled to 10 GHz frequency).}
    \label{phasenoise}
\end{figure}

Approximately 0.25~mW of optical power from the laser oscillator was coupled into a highly linear InGaAs photodiode (Discovery Semiconductor DSC40S) and the resulting electrical signal was filtered, amplified, and analyzed for phase noise and relative intensity noise using a signal source analyzer (Anapico APPH20G).
Figure \ref{phasenoise} shows the power spectral density of the phase noise of the 10-GHz repetition rate in both free-running and locked conditions together with the phase noise of the 1-GHz reference (rescaled to a 10-GHz carrier forcomparison).
It should be noted that in locked condition for offset frequencies lower than 100~Hz the phase noise is limited by the 1-GHz reference whereas for frequencies higher than 5~kHz the phase noise is slightly degraded by the control loop bandwidth (servo bump at 8~kHz).
For offset frequencies higher than 200~kHz, the phase noise reaches a white noise contribution at -147~dBc/Hz, limited by the measurement system noise floor, which is dominated by photodiode shot noise and the instrument noise figure.
The integrated phase noise in the bandwidth from 1~Hz to 1~MHz is 10.2~mrad in locked conditions, corresponding to a timing jitter of the repetition rate of less than 163~fs. 

\begin{figure}[]
    \centering
    \includegraphics[width=\columnwidth]{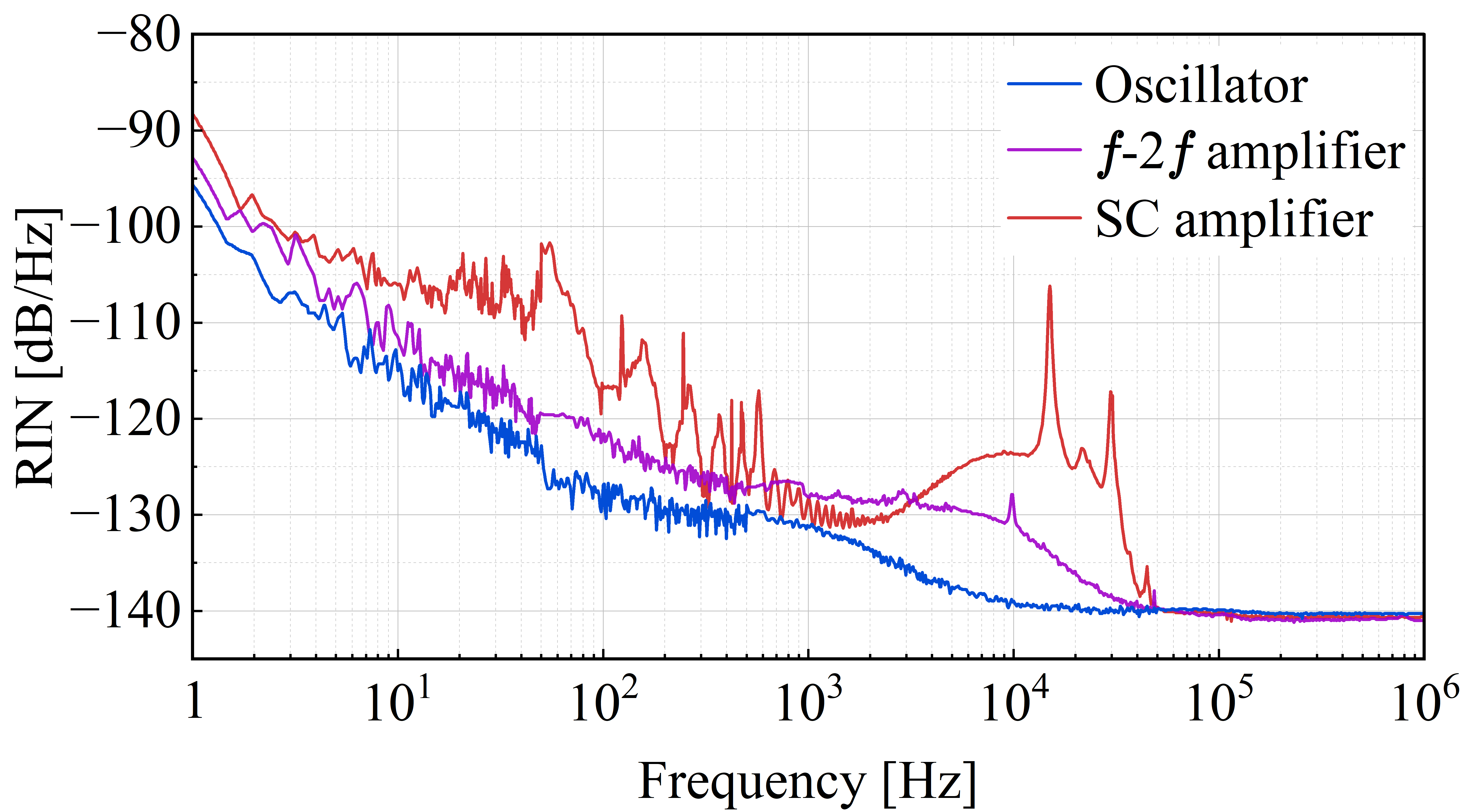}
    \caption{Power spectral density of the relative intensity noise of the 10 GHz oscillator (red trace), compared with those at the optical amplifier outputs:  $f-2f$ branch (blue trace) and SC branch (green trace).}
    \label{RIN}
\end{figure}

Figure \ref{RIN} shows the power spectral densities of the relative intensity noise (RIN) measured at the output of the mode-locked laser and at the output of the CPA stages at the maximum power levels.
Although both amplification stages degrade the intensity noise of the oscillator, for Fourier frequencies higher than 50~kHz the instrument background white frequency noise at $-$140~dB/Hz is reached.
The intensity noise degradation, mainly occuring in the Fourier frequency range from 10~Hz to 50~kHz, is ascribed to the intensity noise of the amplifier pump diodes.
In any case, the RIN remains well below $-$120~dB/Hz even at Fourier frequencies greater than 300~Hz. By numerical integration over the frequency bandwidth from 1~Hz to 1 MHz a RIN better than 0.2~\textperthousand  is obtained for the noisiest amplifier stage (SC branch), whereas for the $f-2f$ branch the integrated RIN is 0.1~\textperthousand. An additional reduction of this noise level can be readily achieved by employing conventional active noise-eater schemes that incorporate external acousto-optic intensity modulators at the outputs of the optical amplifiers \cite{Kalubovilage_2020}. 

\begin{figure}[h]
    \centering
    \includegraphics[width=\columnwidth]{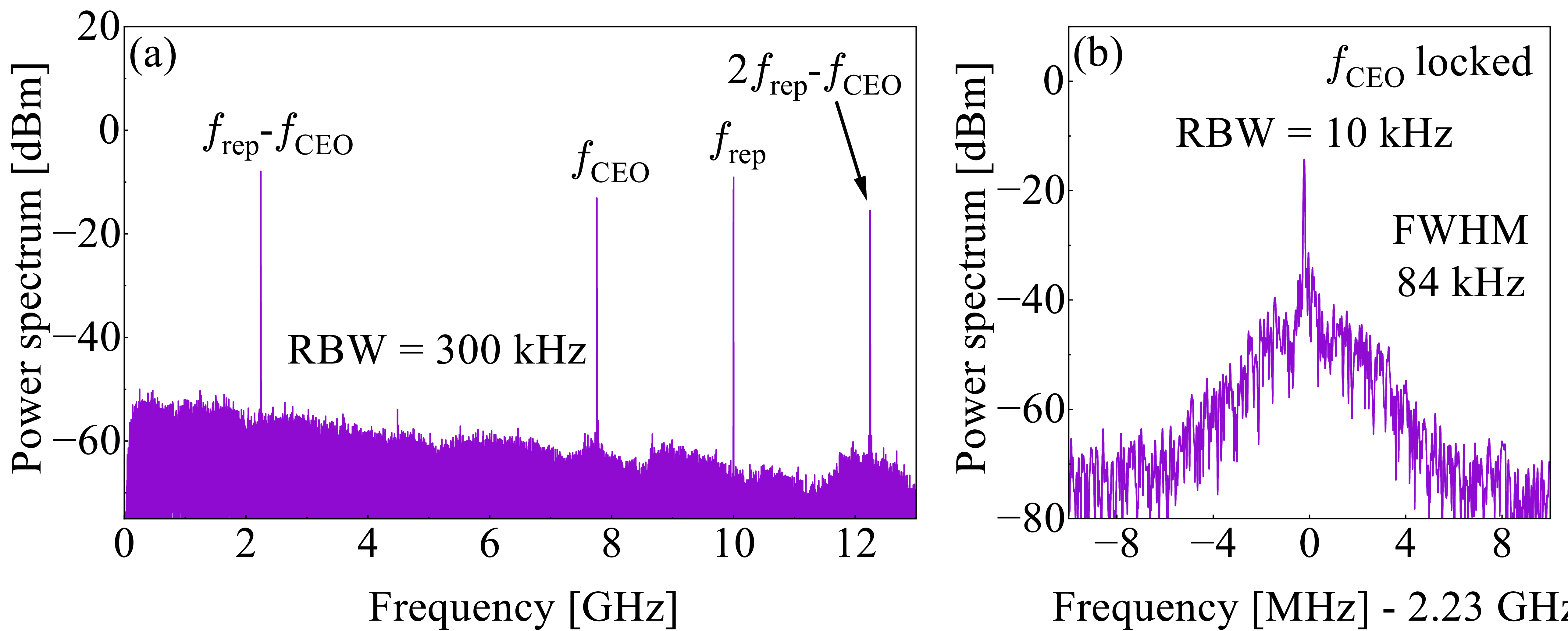}
    \includegraphics[width=\columnwidth]{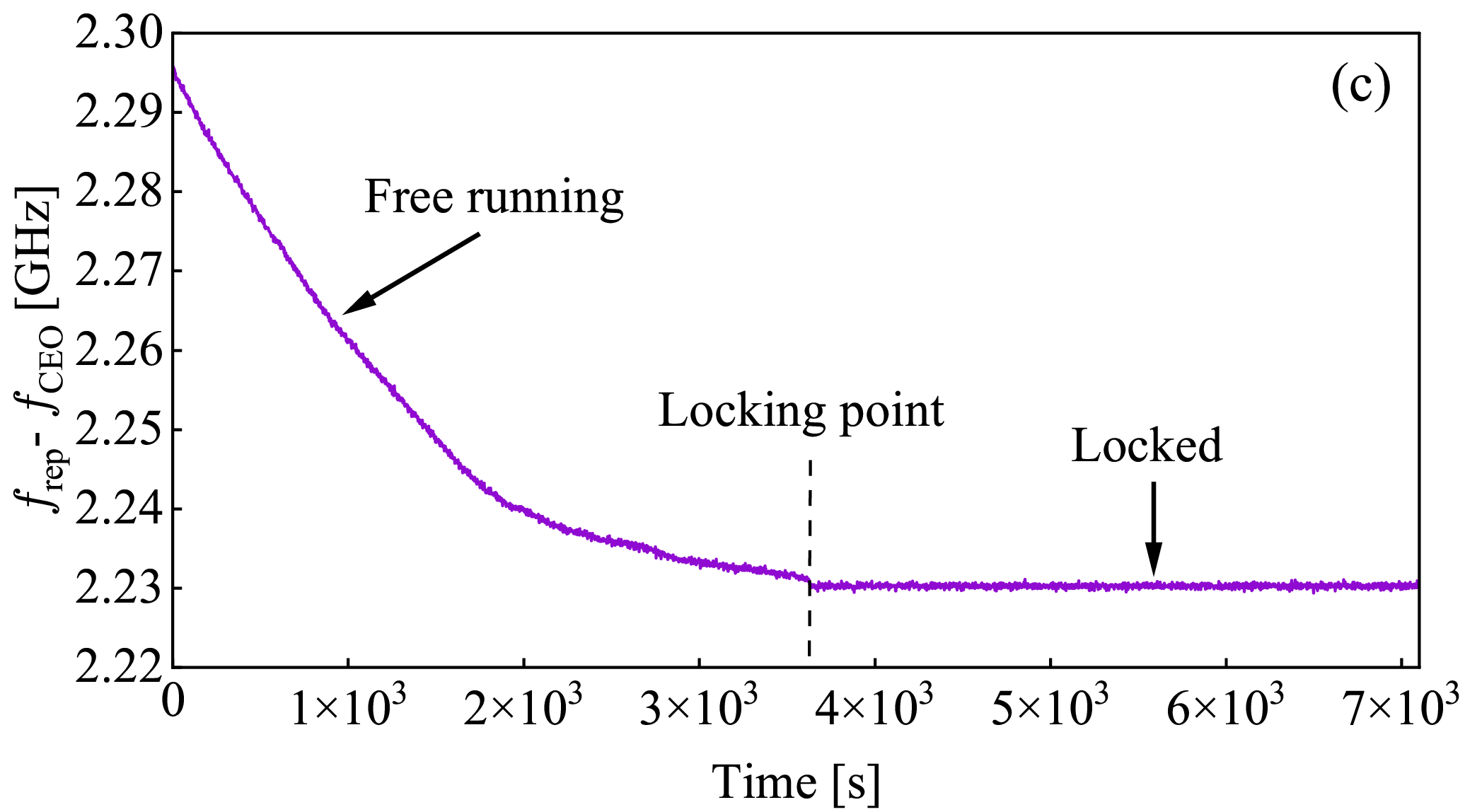}
    \caption{(a) RF spectrum of $f_{CEO}$ from the $f-2f$ interferometer, where three replicas are visible together with the repetition rate. (b) Zoom of the locked CEO, acquired with 10 kHz resolution bandwidth and 20 MHz span. (c) $f_{rep}-f_{CEO}$ as a function of time in both free-running and locked conditions.}
    \label{ceo}
\end{figure}

Figure \ref{ceo} shows the RF spectrum of the photocurrent at the output of fast GaAs photodiode detecting the $f-2f$ beatnote, for an input average power to the tantala nonlinear waveguide of approximately 2~W. 
In particular, in the spectrum of Fig.\ref{ceo}(a) also the pulse repetition rate at 10~GHz is observed together with the aliased carrier envelope offset frequency signals at 2.2~GHz ($f_{rep}- f_{\text{CEO}}$) and 12.2~GHz (2$f_{rep}-f_{\text{CEO}}$). Indeed, by heterodyning a comb tooth with the narrow-linewidth single frequency laser and determining the comb mode order with the aid of a wavemeter (EXFO WA-1500), the $f_{CEO}$ was unambiguously identified at 7.8~GHz. It is worth noting an $f_{CEO}$ signal to noise ratio (SNR) of 50 dB in a 300~kHz resolution bandwidth.

Figure \ref{ceo}(b) reports a zoomed view of the $f_{\text{rep}}-f_{\text{CEO}}$ beatnote, span 20 MHz with a resolution bandwidth of 10~kHz, in locked configuration.
A SNR as large as 60~dB and a FWHM linewidth of 84~kHz are obtained.
We frequency-locked the $(f_{\text{rep}}-f_{\text{CEO}})/4$ frequency, as obtained at the output of a by 4 fast frequency prescaler, against a RF synthesizer, referenceable to GPS, at 550~MHz by using a digital PID servo that controls the current of the pump diode of the 10-GHz laser oscillator. 
Due to a strong control loop bandwidth limitation of the pump diode current actuator ($-$3 dB bandwidth of 7 kHz), and to a nearly white phase noise contribution at $-$55 dBc/Hz in the frequency range from 10 Hz to 100 kHz, a instrument limited phase coherent peak is not observed, indicating that the $f_{\text{CEO}}$ is frequency locked, but not fully phase locked to the RF reference.
If required, full phase locking can be achieved through future upgrades, such as increasing the actuation bandwidth of the pump diode in combination with a large frequency-division factor applied to the $f_{\text{CEO}}$ signal.
Figure~\ref{ceo}(c) reports the $f_{\text{CEO}}$ frequency as a function of the observation time in both free-running and stabilized conditions.
In free-running, the stability of $f_{\text{CEO}}$ is mainly limited by the evolution of environmental conditions (temperature, pressure, and humidity), leading to a peak to peak variation of 80~MHz in a measurement time of approximately 1~h.
The stabilization of$f_{\text{CEO}}$ allows the full referencing of the 10-GHz comb for several hours, with a $\sigma$ deviation of 18.5~kHz for an integration time of 1~ms, over 1 hour observation time.

By accurate measurements of the $f_{\text{rep}}$ and $f_{\text{CEO}}$ frequencies as a function of the signals applied to the laser’s actuators (the piezoelectric actuator, which controls the cavity length, and the current actuator, which controls the pump diode power), we also determined the fixed points of the frequency comb, according to the elastic tape model \cite{Telle_2002,Newbury_2007}, to be located at 55.82~THz for the current actuator and at 0.950~THz for the piezoelectric actuator.
These results confirm that the laser actuators are nearly orthogonal: modulation of the current predominantly influences $f_{\text{CEO}}$, whereas the PZT primarily governs $f_{\text{rep}}$. 

\begin{figure}[t!]
    \centering
    \includegraphics[width=\columnwidth]{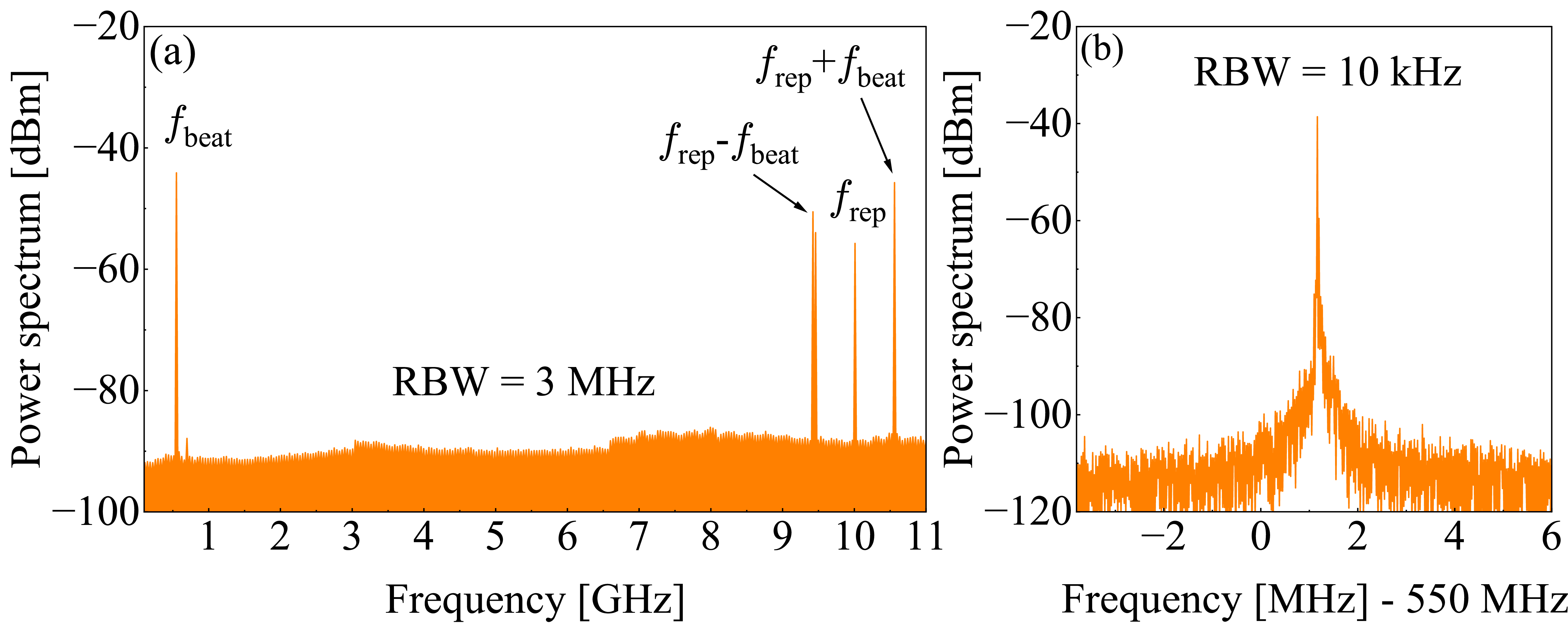}
    \caption{(a) RF spectrum of the beatnote between the ultrastable CW laser and the 10-GHz comb. In the same plot there are three replicas of the beatnote and the repetition rate tone. (b) Zoom of the beat note with the locked comb, acquired with 10 kHz resolution bandwidth and 10 MHz span.}
    \label{beat}
\end{figure}

To directly measure the linewidth of the optical comb tooth at around 1560~nm, a heterodyne beatnote between the 10-GHz Er-doped solid-state comb and a narrow emission linewidth single-frequency laser at 1559.513~nm is detected with a fast InGaAs photodetector (Thorlabs DXM30AF, 30-GHz bandwidth).
Before the InGaAs photodetector, not shown in Fig.~\ref{setup}, a fiber-coupled optical grating filter (JDS Uniphase TB9) with a 0.3~nm bandwidth is used to reduce the detected number of comb modes and increase the SNR of the beatnote.
Figure~\ref{beat}(a) reports the RF spectrum of the InGaAs photocurrent in a span of 11 GHz and a resolution bandwidth of 3~MHz. The beat note signal between the single-frequency laser and the nearest comb mode is detected at 550~MHz with a SNR of 47~dB.
In the same frequency span, apart from the $f_{\text{rep}}$ peak at 10~GHz, the higher order beatnotes at 9.45 and 10.55 GHz ($f_{\text{rep}}\pm f_{\text{beat}}$) have been observed. The FWHM linewidth of the beat note is limited by the spectrum analyzer (Agilent E4445A) to about 10~kHz, as shown in panel (b) of Fig.~\ref{beat}. At this spectral resolution, corresponding to an integration time of 0.1~ms, the beatnote linewidth is determined by that of the frequency comb mode since the contribution from the single-frequency laser linewidth is well below 1~kHz.

\begin{figure}[h!]
\centering
    \includegraphics[width=\columnwidth]{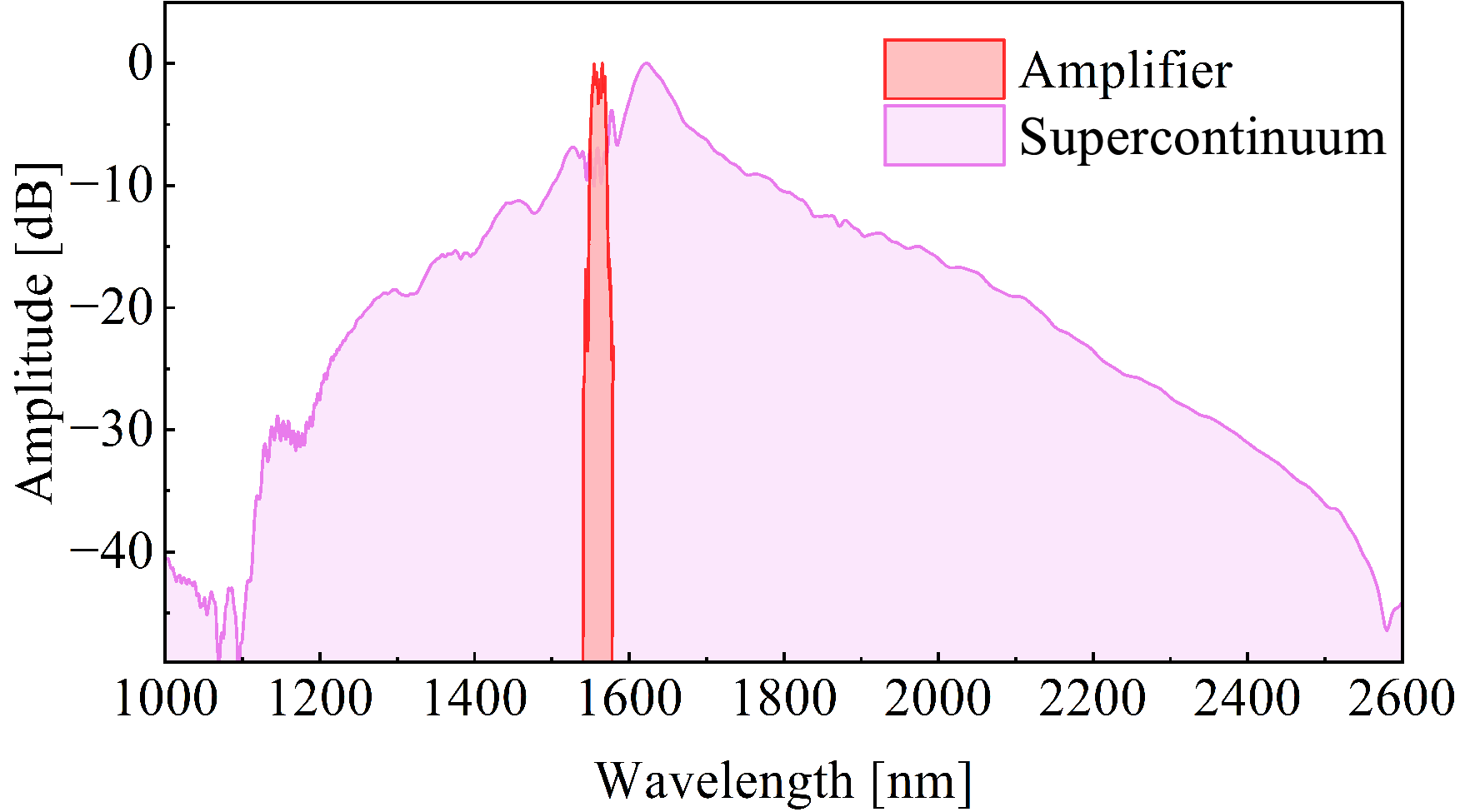}
    \caption{Optical spectrum of the supercontinuum at the system output as measured by a FTIR spectrometer with a frequency resolution of 16~cm$^{-1}$ (480~GHz). The nonlinear tantala waveguide is pumped with \SI{2.2}{\watt} and the output average power is $\sim$800~mW.}
    \label{sc}
    \end{figure}

Figure~\ref{sc} shows the SC spectra generated at the output of the SC tantala nonlinear waveguide together with the spectrum of the amplified laser oscillator.
Increasing the pump power to 2.2 W, corresponding to a pulse peak power of 1 kW, a SC spectrum spanning more than one-octave, from 1150~nm to 2350~nm at -30 dB level, is obtained with an average output power of nearly 800~mW, corresponding to a maximum power per comb tooth exceeding 600~$\upmu$W in the center of the spectrum.
Stable operation of the SC over hours has been observed without degradation of the bandwidth. 

\section{Conclusion}
In summary, we have demonstrated a solid-state optical frequency comb seeded by an Er-doped mode-locked laser at 10 GHz fundamental repetition rate, stabilized both in repetition rate and carrier-envelope offset.
The integrated phase noise of the locked repetition rate in the bandwidth 1 Hz-1 MHz is 10.2 mrad, corresponding to a timing jitter of less than 163 fs.
Such a comb is based on all-PM-coupled fiber technology and commercially available components covering a broad near-infrared spectral range from 1150 to 2350~nm, providing an average power of $\sim$800~mW and a power per comb tooth above 100~$\upmu$W in a spectral span exceeding 200~nm.
The integrated relative intensity noise  (between 1 Hz and 1 MHz) measured at the output of the amplifiers seeding the supercontinuum waveguide is less than 0.2~\textperthousand. An intensity noise lower by a factor 2 is found for the amplifier seeding the $f-2f$ stage, where a $f_{\text{CEO}}$ signal of 50~dB is detected with a 300-kHz resolution bandwidth.
These results establish a robust and accessible platform for high-repetition-rate comb generation, bridging laboratory-grade performance with practical deployment in optical communication, precision timing, and astronomical instrumentation.

\begin{backmatter}
\bmsection{Funding} 
European Union’s NextGenerationEU - Near-Infra-Red Astrocomb for Earth-like exoplanet detections "NIR-Astrocomb" 2022CCWJYL – CUP D53D23002570006 - Grant Assignment Decree No. 962 adopted on 06/30/2023 by the Italian Ministry of Ministry of University and Research (MUR).
European Union’s NextGenerationEU Programme with the STILES Infrastructure [IR0000034, ID ..., CUP C33C22000640006] Strengthening the Italian leadership in ELT and SKA. 
European Union’s NextGenerationEU Programme with the I-PHOQS Infrastructure [IR0000016, ID D2B8D520, CUP B53C22001750006] Integrated infrastructure initiative in PHOtonic and Quantum Sciences. 

\bmsection{Acknowledgment} P. Laporta and F. Leone acknowledge financial support under the National Recovery and Resilience Plan (NRRP), Mission 4, Component 2, Investment 1.1, Call for tender No. 104 published on 2.2.2022 by the Italian Ministry of University and Research (MUR), funded by the European Union – NextGenerationEU - Near-Infra-Red Astrocomb for Earth-like exoplanet detections "NIR-Astrocomb" 2022CCWJYL – CUP D53D23002570006 - Grant Assignment Decree No. 962 adopted on 06/30/2023 by the Italian Ministry of Ministry of University and Research (MUR).
F. Leone, P. Laporta, and G. Galzerano acknowledge financial support by the European Union’s NextGenerationEU Programme with the STILES Infrastructure [IR0000034, ID XXXX, CUP C33C22000640006] Strengthening the Italian leadership in ELT and SKA.  
G. Galzerano acknowledge financial support by the European Union’s NextGenerationEU Programme with the I-PHOQS Infrastructure [IR0000016, ID D2B8D520, CUP B53C22001750006] Integrated infrastructure initiative in PHOtonic and Quantum Sciences. 

\bmsection{Disclosures} A. Pertoldi, B. Rudin, O. Kara, and F. Emaury disclose association with Menhir Photonics, which aims to commercialize ultrafast 10-GHz laser.

\bmsection{Data Availability}
According to the open data policies of the Physics Department of Politecnico di Milano, all data underlying the results presented in this paper are publicly available in \cite{}.

\end{backmatter}

\bibliography{References}

\bibliographyfullrefs{References}


\end{document}